\documentclass[epsfig,12pt]{article}
\usepackage{graphicx,epsfig}
\def\gev{\,{\rm GeV}}

\def\to{\rightarrow}

\begin{document}
\begin{titlepage}
\rightline{\vbox{\halign{&#\hfil\cr &KEK-TH-965
\cr &hep-ph/0406121\cr
&2004\cr}}}
\vskip .5in

\begin{center}
\Large{{\bf Leading power SCET analysis of $e^+ e^- \to J/\psi g g$}}
\vskip 1cm

\normalsize {Zhi-Hai Lin$^{1}$ and Guohuai Zhu$^{2}$ \\
\vskip .5cm
 {  Theory Group, KEK, Tsukuba, Ibaraki 305-0801, Japan}
  \footnote{Email: linzh@post.kek.jp}\\
\vskip .3cm
 {  Centre de Physique Theorique, Centre National de la
Recherche Scientifique, UMR 7644, Ecole Polytechnique, 91128 Palaiseau
Cedex, France} \footnote{Email: Guohuai.Zhu@cpht.polytechnique.fr} }

\vskip 1cm
\end{center}
\begin{abstract}
Recently, Belle and BaBar Collaborations observed surprising suppression
in the endpoint $J/\psi$ spectrum, which stimulates us to examine the
endpoint behaviors of the $e^+ e^- \to J/\psi gg$ production. We calculate
the $J/\psi$ momentum and angular distributions for this process within
the framework of the soft-collinear effective theory (SCET). The decreasing
spectrum in the endpoint region is obtained by summing the Sudakov logarithms.
We also find a large discrepancy between the NRQCD and SCET spectrum in
the endpoint region even before the large logarithms are summed,
which is probably due to the fact that only the scalar structure
of the two-gluon system is picked out in the leading power expansion.
A comparison with the process $\Upsilon \to \gamma gg$ is made.
\end{abstract}
\end{titlepage}

\section{Introduction}
Heavy quarkonium system plays an important role in the development
of quantum chromodynamics (QCD). The scale of the heavy quark mass
guarantees the applicability of perturbative QCD, meanwhile the
nonperturbative physics presents itself through hadronization
effects. In the past few years, one of developments in heavy
quarkonium physics, called nonrelativistic quantum chromodynamics
(NRQCD)~\cite{NRQCD} which generalizes and improves the conventional
color-singlet model (CSM), has provided a successful explanation of
the surprising excesses of $J/\psi$ and $\psi'$ productions at the
Tevatron \cite{CDF} by introducing color octet contributions.

NRQCD factorization should be further examined in other collider
facilities, in particular, $e^+ e^-$ colliders which provide a clean
testing ground. SLAC and KEK $e^+ e^-$ B factories are now running
at or below the $\Upsilon(4s)$ resonance. At this energy, it was
expected in NRQCD that the inclusive $J/\psi gg$ process should be
dominant \cite{cho,Lin} and in the upper endpoint region of the
$J/\psi$ momentum spectrum, there may exist a sharp peak as a clean
signal of the color-octet $c \bar{c} g$ contribution
\cite{braaten-chen}.

Recently, BaBar \cite{babar} and Belle \cite{belle1,belle2} Collaborations
published their measurements on prompt $J/\psi$ productions in $e^+ e^-$
collision at center-of-mass (c.m.) energy $\sqrt{s}=10.58~\gev$. It is
really surprising to observe that, according to Belle's data \cite{belle1,
belle2}, it is the $J/\psi c {\bar c}$ process that dominates the
inclusive $J/\psi$ production at B factories
\begin{equation}\label{rate2}
\sigma(e^+ e^- \to J/\psi c {\bar c})/\sigma(e^+ e^- \to J/\psi X) =
0.67\pm 0.12,
\end{equation}
while the momentum distribution of the inclusive $J/\psi$ production
shows a suppression, instead of an (expected) enhancement, in the upper
endpoint region.

For the unexpected $J/\psi c \bar{c}$ dominance, it is argued in Ref.
\cite{Lin} that a large renormalization $K$ factor might be the answer.
Recent investigation \cite{FLM03} also reveals that the color-octet
contribution to $J/\psi$ spectrum can be broadened significantly by the
large perturbative corrections and enhanced nonperturbative effects so as
not to conflict with the surprising suppression in the endpoint region
observed by BaBar and Belle.
However a leading-order NRQCD calculation shows that, in the endpoint
region, the color-singlet $J/\psi gg$ contribution is not small at all,
which seems to be still in contradiction with the experimental
observations. In this work, we are stimulated to investigate
the endpoint behaviors of the $e^+ e^- \to \gamma^\ast \to J/\psi g g$
production.

We note that, at the amplitude level, $\gamma^\ast \to J/\psi gg$ is
very similar to the decay $\Upsilon \to \gamma g g$. It has been
known several years ago that, at the endpoint of the photon spectrum
in radiative $\Upsilon$ decays, NRQCD is not applicable due to the
breakdown of both the perturbative expansion and the operator
product expansion (OPE) \cite{pe-ope}. The same arguments should
also apply for the case of the $J/\psi$ production. This is because
 NRQCD only contains soft degrees of freedom at low energy, but at
the endpoint of the photon and/or $J/\psi$ spectrum, the gluon jet
should be almost collinear. To fix this problem, Fleming {\it et
al.} proposed a combination of NRQCD for the heavy degrees of
freedom  and the soft-collinear effective theory (SCET) \cite{SCET}
for the light degrees of freedom. With this method, the radiative
$\Upsilon$ decays were investigated in a series of papers
\cite{bauer, fleming, soto} which show an improved agreement with
the CLEO data \cite{CLEO}. Lately the same method was applied to the
color-octet contribution to the inclusive $J/\psi$ production $e^+
e^- \to J/\psi+X$ \cite{FLM03}. By the use of the resummation of
Sudakov logarithms and the nonperturbative shape functions, the
color-octet $J/\psi$ spectrum, which is a sharp peak at maximal
energy in leading order calculations, could be significantly
broadened and shifted to lower energies. This therefore would
resolve the discrepancy between the color-octet $J/\psi$ production
and the experimental observations. According to the spirit of the
Sudakov suppression in the endpoint region, the authors in Ref.
\cite{Lin} adopted a phenomenological approach to obtain an
appropriate endpoint spectrum for the $J/\psi gg$ process instead of
performing a complete calculation in SCET.

In this paper, we shall follow the same way of Ref.~\cite{fleming,FLM03},
namely SCET combined with NRQCD, to examine the endpoint behavior of
the color-singlet $J/\psi g g$ mechanism.

\section{Leading order SCET calculation}
Several scales are involved in this process: the center-of-mass energy
$\sqrt{s}$, the $J/\psi$ mass $M_{\psi}$, and the non-perturbative
QCD scale $\Lambda_{QCD}$. In this paper we will only consider the
case where the ratio $M_{\psi}/\sqrt{s}$ is kept finite in the limit
of infinite $\sqrt{s}$. In this point of view, $J/\psi$ can be taken as a
heavy particle. In the kinematic endpoint region
of $J/\psi$ spectra, the failure of NRQCD factorization and the relevance
of SCET has been explained clearly in Refs. \cite{FLM03,fleming}.
In brief, the hadronic jet recoiling against $J/\psi$ is not highly
virtual, $m_X \sim \sqrt{\sqrt{s} \Lambda_{QCD}}$, compared with its
large momentum of order $\sqrt{s}$.
This results in the OPE breaking down, and therefore a new effective theory,
the so-called SCET, is developed by including collinear degrees of freedom.

In SCET, it is convenient to write a momentum in light-cone coordinates.
Working in the $e^+ e^-$ c.m. frame, we define the incoming electron and
positron moving along light-cone directions $n_e^\mu=(1,0,0,-1)$ and
${\bar n}_e^\mu=(1,0,0,1)$.
The produced $J/\psi$ meson is chosen to move in the $x-z$ plane with
momentum $P_{\psi}^\mu=M v^\mu=(E, |\vec P| \sin \theta, 0, |\vec P|
\cos \theta)$ (M is $J/\psi$ mass),
and hence the light-cone vectors for two gluons can be defined as
$n^\mu=(1,-\sin \theta, 0, -\cos \theta)$
and ${\bar n}^\mu=(1,\sin \theta, 0, \cos \theta)$. Throughout this
paper, we adopt a dimensionless variable $z= |\vec P_\psi|/P_\psi^{max}$,
where $P_\psi^{max}$ denotes the maximum value of the $J/\psi$ momentum,
namely $P_{\psi}^{max}=\sqrt{s}(1-r)/2\sim 4.9 \gev$. Here $r=M^2/s\sim
0.08$. The $J/\psi$ velocity $v$ can be expressed as
\begin{equation}\label{v}
v^\mu=(~v_0~,~|{\vec v}| \sin \theta~,~0~,~|{\vec v}| \cos \theta~)
 =(~\sqrt{\frac{(1-r)^2}{4r}z^2+1}~,~\frac{1-r}{2\sqrt{r}}z \sin \theta
   ~,~0~,~\frac{1-r}{2\sqrt{r}}z \cos \theta~).
\end{equation}
For the process $e^+ e^- \to \gamma^*\to J/\psi X$, the hadronic jet
has the momentum $p_X^\mu=l^\mu-M v^\mu-k^\mu$,
where $l^\mu=(\sqrt{s},0,0,0)$ is the momentum of the virtual photon and
$k^\mu$ is the residual momentum of the $c \bar{c}$ pair within $J/\psi$.
In the endpoint region, since the hadronic jet is collinear along the
light-cone direction $n^\mu$, we can write $p_X \sim \sqrt{s}
(1,\lambda^2,\lambda)$ in the $n-\bar{n}$ light-cone coordinate.
When $E_\psi^{max}-E_\psi \sim \Lambda_{QCD}$, $p_X^2$ is of order
$2\sqrt{s} (E_\psi^{max}-E_\psi) \sim 2\sqrt{s} \Lambda_{QCD}$ which implies
NRQCD factorization breaks down in this kinematic region. Therefore SCET
becomes relevant in the endpoint region $1-z \sim \Lambda_{QCD}/M \sim
v^2$, and correspondingly the expansion parameter $\lambda$ is of order
$\sqrt{1-z}$ in this process.

Before going into details, it is helpful to notice the similarity between
$e^+ e^- \to \gamma^* \to J/\psi gg$ and $\Upsilon \to \gamma g g$.
In fact the cross
section of $J/\psi gg$ production can be related to the ``decay width"
of the transversely polarized virtual photon $\gamma^\ast$ as follows
\cite{Keung}
\begin{equation}
d\sigma(e^+ e^- \to J/\psi gg)=4\pi \alpha s^{-3/2}
d\Gamma(\gamma^\ast \to J/\psi gg)~,
\end{equation}
where the polarization vector of the virtual photon satisfies the
following equation
\begin{equation}\label{pola}
\epsilon^\mu \epsilon^{\ast \nu}=-g^{\mu \nu}+\frac{n_e^\mu {\bar n}_e^\nu
     +n_e^\nu {\bar n}_e^\mu}{2}\equiv -g^{\mu
      \nu}_{\perp e}~.
\end{equation}
It is then clear that, at the amplitude level, the effective operator of
$\gamma^\ast \to J/\psi gg$ should be formally the same as that of
$\Upsilon \to \gamma g g$. Therefore the proof of SCET factorization for
the former process is almost the same as that of the latter one, which has
been elaborated in Ref. \cite{fleming}. All of our following calculations
will be in parallel with those for $\Upsilon \to \gamma gg$ in
Ref. \cite{fleming}.

To proceed, we shall first match from NRQCD onto SCET. Considering
the gauge and reparametrization invariance, the leading SCET
color-singlet ${}^3 S_1$ operator is given by \cite{fleming}
\footnote[1]{Here `leading' means that the leading-order power
expansion in terms of the small parameter $\lambda$ in SCET.}
\begin{eqnarray}\label{operator}
{\cal O}(1, {}^3 S_1)&=&\mbox{\boldmath $\psi$}^\dag_{\mathbf{p}} \Lambda
\cdot \mbox{\boldmath $\sigma$}^\delta \chi_{-\mathbf p}
{\rm Tr}\{ B_\perp^\alpha
\Gamma_{\alpha \beta \delta \mu}^{(1, {}^3 S_1)}(-n \cdot v {\bar{\cal
P}}, -n \cdot v {\bar{\cal P}}^\dag) B_\perp^\beta\} \nonumber \\
&=&\mbox{\boldmath $\psi$}^\dag_{\mathbf{p}} \Lambda
\cdot \mbox{\boldmath $\sigma$}^\delta \chi_{-\mathbf p}
{\rm Tr}\{ B_\perp^\alpha
\Gamma_{\alpha \beta \delta \mu}^{(1, {}^3 S_1)}(\frac{M(1-r)}{r}, -n
\cdot v {\cal P}_-) B_\perp^\beta\}~,
\end{eqnarray}
where $\psi_{\mathbf p}$ and $\mbox{\boldmath $\chi$}_{- \mathbf{p}}$
are the heavy quark and antiquark fields from NRQCD, and $B_\perp$ is the
leading piece of the collinear-gauge invariant gluon field strength
\cite{fleming}. The
operator ${\bar{\cal P}}$(${\bar{\cal P}}^\dag$) projects out the large
light-cone momentum components of the collinear fields to the right(left).
The second line of the above equation is obtained by using the identity
$ B_\perp^\alpha n \cdot v ({\bar{\cal P}}+{\bar{\cal P}}^\dag)
B_\perp^\beta= -M(1-r)/r B_\perp^\alpha B_\perp^\beta $ and the definition
${\cal P}_-=\bar{\cal P}-\bar{\cal P}^\dag$.
From the matching shown in Fig. 1, we obtain the coefficient
\begin{equation} \label{coefficient}
\Gamma_{\alpha \beta \delta \mu}^{(1, {}^3 S_1)}
(\frac{M(1-r)}{r}, n \cdot v \bar{n} \cdot q_- )=
-\frac{4g_s^2 e e_c}{\sqrt{6}} \frac{r}{M(1-r)} g^\perp_{\alpha \beta}
(g_{\mu \delta}+\frac{1-r}{2 r}n_\mu \bar{n}_\delta)~,
\end{equation}
where $g^\perp_{\alpha \beta}=g_{\alpha \beta}-(n_\alpha \bar{n}_\beta
 +n_\beta \bar{n}_\alpha)/2$,
$\bar{n} \cdot q_-=\bar{n} \cdot q -
\bar{n} \cdot q'$. $q$ and $q'$ are the momenta of two gluons.

\begin{figure}
\centerline{\epsfysize 2.0 truein \epsfbox{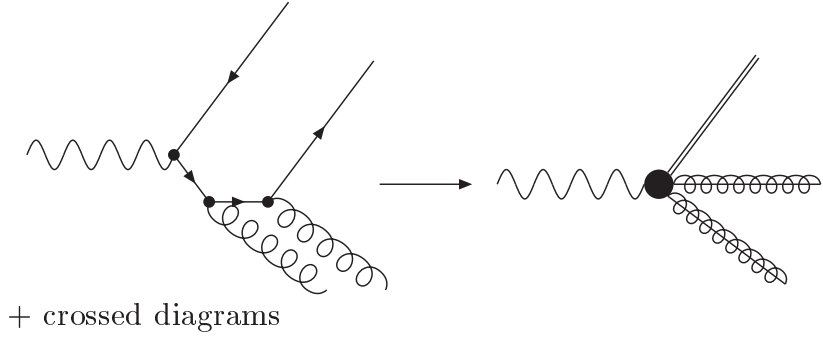}}
\caption{Matching the amplitude for $\gamma^\ast \to
J/\psi gg$ process in QCD and SCET.}
\end{figure}

According to the optical theorem, the $J/\psi$ momentum spectrum and angular
distribution can be expressed as
\begin{equation}
\frac{d \Gamma(\gamma^\ast \to J/\psi gg)}{dzd\cos\theta}
= \frac{(P_{\psi}^{max})^3 z^2 }{8 \pi^2 \sqrt{s}
 \sqrt{M^2+(P_{\psi}^{max})^2 z^2} }
{\rm Im} T(z,\theta)
\end{equation}
where the forward scattering amplitude is
\begin{equation}\label{forward}
T(z,\theta)=-i \int d^4 x e^{-i l \cdot x} \sum_X \langle 0 \vert
J_v^\dag (x) \vert J/\psi + X \rangle \langle J/\psi + X \vert J_\mu(0)
\vert 0 \rangle g^{\mu \nu}_{\perp e}~.
\end{equation}
In SCET, the following factorization formula can be proved in
the endpoint region
\begin{eqnarray}\label{fact}
{\rm Im} T(z,\theta)&=&\sum_\omega H(\frac{M(1-r)}{r},
\omega,z, \theta, \mu)  \nonumber \\
& &\hspace*{0.2cm} \times \int dk^+ S(k^+, \mu) {\rm Im} J_\omega(k^+
+\sqrt{s}-P_{\psi}^{max}z-
\sqrt{M^2+(P_{\psi}^{max})^2 z^2}, \mu)~,
\end{eqnarray}
where $H$, $S$ and $J_\omega$ are the hard function, ultrasoft function and
jet function, respectively. In order to obtain the above formula, we
match the QCD current $J_\mu$ in Eq. (\ref{forward}) to the leading SCET
color-singlet operator Eq. (\ref{operator})
\begin{equation}
J_\mu(x)=\sum_\omega e^{-i(Mv-\bar{\cal P}(n/2))\cdot x} i
\Gamma_{\alpha \beta \delta \mu}^{(1, {}^3 S_1)}(\omega)
\mbox{\boldmath $\psi$}^\dag_{\mathbf{p}} \Lambda \cdot
    \mbox{\boldmath $\sigma$}^\delta \chi_{-\mathbf p}
  {\rm Tr}\{B_\perp^\alpha \delta_{\omega,{\cal P}_-}B_\perp^\beta \}~,
\end{equation}
where the operator $\bar{\cal P}$ in the phase factor will sum the
label momentum of the two collinear fields $B_\perp$ and thus can be
replaced by $-\sqrt{s}(1-r)$. The matching coefficient
$\Gamma_{\alpha \beta \delta \mu}^{(1, {}^3 S_1)}(\omega)$ is given
in Eq. (\ref{coefficient}). Since collinear fields in SCET are
decoupled from ultrasoft gluons by field redefinition \cite{SCET}
and $J/\psi$ meson has no collinear freedom, the forward scattering
amplitude in Eq. (\ref{forward}) can then be factorized by
separating the heavy quark fields into ultrasoft functions and the
collinear gluon fields into jet functions. Specifically, the jet
function is defined from the vacuum matrix element of the collinear
fields, which is exactly the same as that of the color-singlet
radiative $\Upsilon$ decay \cite{fleming}
\begin{eqnarray}\label{jet}
&&\langle 0 \vert T~{\rm Tr}[B_\perp^{0\alpha} \delta_{\omega,{\cal P}_-}B_\perp^{0\beta}](x)
 {\rm Tr}[B_\perp^{0\alpha'} \delta_{\omega',{\cal P}_-}B_\perp^{0\beta'}](0) \vert 0 \rangle
\nonumber \\ && \hspace*{0.5cm}
\equiv \frac{i}{2}(g_\perp^{\alpha \alpha'}g_\perp^{\beta \beta'}+g_\perp^{\alpha \beta'}
 g_\perp^{\alpha' \beta} ) \delta_{\omega,\omega'} \int \frac{d^4 k}{(2\pi)^4}
 e^{-ik\cdot x} J_\omega (k^+,\mu)~,
\end{eqnarray}
where $B_\perp^0$ is the redefinition of the collinear field to decouple
from the ultrasoft gluons. To calculate the jet function,
one may directly evaluate the vacuum matrix element of the collinear
fields, which is the left-hand side of Eq. (\ref{jet}).
Actually the jet function which is independent of the heavy quark fields,
should be the same for both $\Upsilon \to \gamma gg$ and $\gamma^\ast \to
J/\psi gg$  processes, so it can be obtained directly from
Ref. \cite{fleming}. For our purpose, only the imaginary part of
the jet function is relevant, and at the lowest order in $\alpha_s$, it is
\begin{equation}
{\rm Im}J_\omega(k^+, \mu)=\frac{1}{8\pi} \Theta (k^+)\int_{-1}^{1} d\xi
\delta_{\omega, \sqrt{s}(1-r)\xi}~.
\end{equation}
Following Ref. \cite{fleming}, the ultrasoft
function for this process can be written as
\begin{eqnarray}
S(k^+, \mu)&=&\int \frac{dx^-}{4 \pi} e^{-(i/2)k^+ x^-}
 \langle 0 \vert \mbox{\boldmath $\chi$}^\dag_{-\mathbf{p}}
    \mbox{\boldmath $\sigma$}_i \psi_{\mathbf p}(x^-) a^+_\psi a_\psi
  \mbox{\boldmath $\psi$}^\dag_{\mathbf{p}}
    \mbox{\boldmath $\sigma$}_i \chi_{-\mathbf p} (0)
      \vert 0 \rangle \nonumber \\
 &=&\langle 0 \vert \mbox{\boldmath $\chi$}^\dag_{-\mathbf{p}}
    \mbox{\boldmath $\sigma$}_i \psi_{\mathbf p} a^+_\psi a_\psi
  \delta (in \cdot \partial - k^+ )
  \mbox{\boldmath $\psi$}^\dag_{\mathbf{p}}
    \mbox{\boldmath $\sigma$}_i \chi_{-\mathbf p}
      \vert 0 \rangle~,
\end{eqnarray}
while the leading order hard function is computed as
\begin{eqnarray}
H(\omega, z,\theta, \mu)&=&\frac{2}{3}\left (
 \frac{4g_s^2 e e_c r}{\sqrt{6}M(1-r)} \right )^2 g_{\perp e}^{\mu \nu}
 \left ( g_{\mu \delta}+\frac{1-r}{2 r}n_\mu \bar{n}_\delta  \right )
 \left ( g_{\nu \lambda}+\frac{1-r}{2 r}n_\nu \bar{n}_\lambda \right )
 (g^{\delta \lambda}-v^\delta v^\lambda ) \nonumber \\
&=&\frac{32 \pi^2}{3} \left (
 \frac{4 \alpha_s e e_c r}{\sqrt{6}M(1-r)} \right )^2 F(z,\theta)~,
\end{eqnarray}
where the explicit expression for $F(z,\theta)$ is
\begin{equation}\label{f}
F(z,\theta)=2- \sin^2 \theta + \frac{\sin^2 \theta}{4 r^2}
[ (1+r)v_0-(1-r)|{\vec v}| ]^2~.
\end{equation}
Here $v$ is the $J/\psi$ velocity given in Eq. (\ref{v}).

With these functions in hand, we obtain the explicit form for
the imaginary part of the forward scattering amplitude in Eq. (\ref{fact})
as follows
\begin{eqnarray} \label{ImT}
{\rm Im} T(z, \theta)&=&\Theta(\sqrt{s}-P_\psi-E_\psi)
\frac{16 \pi}{3 M}\left (
 \frac{4 \alpha_s e e_c r}{\sqrt{6}M(1-r)} \right )^2
F(z,\theta)
\langle 0 \vert \mbox{\boldmath $\chi$}^\dag_{-\mathbf{p}}
    \mbox{\boldmath $\sigma$}_i \psi_{\mathbf p} a^+_\psi a_\psi
  \mbox{\boldmath $\psi$}^\dag_{\mathbf{p}}
    \mbox{\boldmath $\sigma$}_i \chi_{-\mathbf p}
      \vert 0 \rangle \nonumber \\
&=&\Theta(\sqrt{s}-P_\psi-E_\psi)
\frac{8 N_c |R(0)|^2}{3 M}\left (
 \frac{4 \alpha_s e e_c r}{\sqrt{6}M(1-r)} \right )^2
F(z,\theta)~,
\end{eqnarray}
where $R(0)$ denotes the radial wave function of $J/\psi$ at the
origin. Using the above equation, we arrive at the differential
cross section in the
tree-level SCET calculation \footnote[2]{
'The tree-level SCET calculation' here is referred to the
leading power calculation in SCET before the resummation over large logarithms.
}
\begin{eqnarray}\label{treeSCET}
\frac{d\sigma_{\rm trSCET}}{dzd\cos\theta}=
\Theta(\sqrt{s}-P_\psi-E_\psi)
\frac{32(\alpha_s \alpha e_c)^2 N_c}{9 s}
\frac{|R(0)|^2}{M^3} \frac{r^2(1-r)z^2}{\sqrt{4r+(1-r)^2 z^2}}
F(z,\theta)~.
\end{eqnarray}

Since OPE breaks down and large logarithms arise as $z$ approaches
to 1, resummation over large logarithms is indispensable before
comparing to the experimental observations. In SCET, these
logarithms can be resummed using renormalization group equations
(RGE). To do this, one has to first calculate the anomalous
dimension of the effective operator (Eq. (\ref{operator})).
Fortunately, this effective operator is formally the same as that
which appears in the color-singlet radiative $\Upsilon$ decays,
therefore we can directly read the anomalous dimension from Ref.
\cite{fleming}
\begin{equation}
\gamma (\eta)=\frac{2}{\beta_0} \left \{ C_A \left [
\frac{11}{6}+(\eta^2+(1-\eta)^2)\left( \frac{\ln \eta}{1-\eta}
+\frac{\ln (1-\eta)}{\eta} \right) \right ] -\frac{n_f}{3} \right \}~.
\end{equation}
With this anomalous dimension, one can then resum the large
logarithms using RGE from the matching (hard) scale to the collinear
scale. The collinear scale should roughly be the invariant mass of
the jet, namely $\mu_c(z)=\sqrt{2\sqrt{s}(E_{\psi}^{
max}-E_\psi(z))}$. However there is no obvious clue what the
matching scale should be. In NRQCD calculations, this scale is often
chosen at quarkonium mass $M$, but in SCET it is found that, at
least the hard scale for color-octet $J/\psi$ production should be
about $-n \cdot v \bar{{\cal P}}=M(1-r)/r$ \cite{FLM03}, according
to the logarithm that appears in the anomalous dimension
calculations. As we know that, there is no large logarithm far from
the endpoint region, which means that the collinear scale, which is
of the order of $\sqrt{s}$ for small $z$, should be comparable to
the hard scale. Therefore in the following, we will naively choose
the hard scale as $\mu_h=\sqrt{s}(1-r)$, which is simply the large
light-cone component of the gluon jet momentum. Finally, the result
for the resummed differential cross section is
\begin{eqnarray}\label{resum}
\frac{d\sigma_{\mbox{\small resum}}}{dzd\cos\theta}&=&
\Theta(\sqrt{s}-P_\psi-E_\psi) \frac{32(\alpha_s(\mu_h) \alpha
e_c)^2 N_c}{9 s} \frac{|R(0)|^2}{M^3}
\frac{r^2(1-r)z^2}{\sqrt{4r+(1-r)^2 z^2}}
\nonumber \\
&& \times F(z,\theta) \int_0^1 d\eta \left (
\frac{\alpha_s(\mu_c(z))}{\alpha_s(\mu_h)} \right )^{2
\gamma(\eta)}~.
\end{eqnarray}

\section{Results and Discussions}

It is understood that SCET is only valid at the large $z$ region,
while NRQCD should be fine in the small and medium $z$ region.
Therefore in order to obtain a formula which can describe the
$J/\psi$ production in the whole kinematic region, one shall
interpolate smoothly between the NRQCD and resummed SCET results.
Here we propose an interpolating formula
\begin{eqnarray}\label{int}
\frac{d\sigma_{\rm int}}{dz d\cos\theta}=
(1-z)\frac{d\sigma_{\rm NRQCD}}{dz d\cos\theta}
+z\frac{d\sigma_{\rm resum}}{dz d\cos\theta}~.
\end{eqnarray}
Obviously the NRQCD contribution vanishes in the limit of $z=1$, and
only the resummed contribution survives, while at the small $z$, the
NRQCD contribution dominates. In addition, if one does not do any
expansion and resummation in SCET, $\sigma_{\rm resum}$ should be
replaced by $\sigma_{\rm NRQCD}$ and hence Eq.~(\ref{int}) will reproduce
the NRQCD result.

The differential cross section for the $J/\psi gg$ production are restricted
by unitarity, parity, and angular momentum considerations. Its polar angle
dependence can be parametrized into the form \cite{cho}
\begin{eqnarray}\label{cos}
\frac{d \sigma}{dz d\cos\theta}=S(z)[1+\alpha(z)\cos^2\theta],
\end{eqnarray}
where the angular coefficient $\alpha(z)$ is generally limited in
the interval $-1 \leq \alpha(z) \leq 1$. This general form has
been confirmed directly by the calculations in the framework of NRQCD
\cite{Keung, cho, Lin}. From Eqs. (\ref{f}), (\ref{treeSCET}) and
(\ref{resum}), it is easy to find that the tree-level and resummed SCET
results also keep the form of Eq. (\ref{cos}), and furthermore,
have the same coefficient $\alpha(z)$. As a natural result, the
interpolated resummed cross section in Eq. (\ref{int}) follows the same
behavior.

In our numerical estimation we use $\sqrt{s}=10.58\gev$ and
$M=2m_c=3.0\gev$. For simplicity, we also normalize the cross
section to a dimensionless quantity by a factor
$R=(128/3)\alpha_s(\mu_h)^2 \alpha^2 e_c^2 M |R(0)|^2/s^3$.

\begin{figure}
\centerline{\epsfysize 2.0 truein \epsfbox{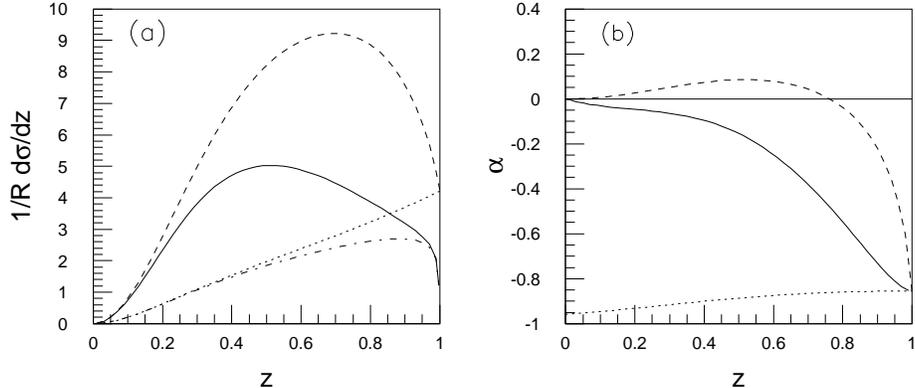}}
\caption{The momentum distribution (a) and angular coefficient $\alpha(z)$
(b) for $e^+ e^- \to J/\psi gg$ process. The dashed, dotted and dot-dashed
curves correspond to the NRQCD, tree-level SCET and resummed SCET
calculations, respectively. The solid curve is for the interpolated
resummed results.}
\end{figure}

In Fig. 2(a), we show the momentum distribution of the process
$e^+ e^- \to J/\psi gg$. The dashed, dotted, dot-dashed and solid curves
correspond to the NRQCD, tree-level SCET, resummed SCET calculations
and the interpolated resummed result, respectively. The NRQCD result
is taken from Ref. \cite{Keung}, while the tree-level SCET, resummed
SCET and interpolated resummed results are obtained by integrating
over the polar angle $\cos\theta$ in Eqs. (\ref{treeSCET}),
(\ref{resum}) and (\ref{int}). As a cross-check of Eq.
(\ref{treeSCET}), one can see that the tree-level SCET result
coincides with the NRQCD one in the limit $z \to 1$. Comparing to
the NRQCD calculation, the interpolated resummed momentum
distribution is suppressed significantly not only in the large $z$
region but also in the medium $z$ region. For example, at $z=0.9$
the ratio of the interpolated resummed cross section and the NRQCD
cross section is about 0.4, while at $z=0.5$ the ratio is still 0.6
which is not quite close to unit. However, the large suppression
might be overestimated. This is because of the large discrepancy
between the NRQCD and tree-level SCET results. Although at $z=1$,
the NRQCD and tree-level SCET results are exactly the same, which is
guaranteed by the matching procedure, the tree-level SCET spectrum
deviates very quickly from the NRQCD one as $z$ departs from one.
For instance, at $z=0.9$, the tree-level SCET cross section is only
half of the NRQCD one. This indicates that the tree-level SCET
calculation may not be a good expansion of the NRQCD calculation
even in the large $z$ region. The resummed SCET cross section is
entirely based on the tree-level SCET calculation, as shown in Eqs.
(\ref{treeSCET}) and (\ref{resum}), and therefore the
over-suppression occurs after interpolating between the NRQCD and
the resummed SCET contributions.

The discrepancy between the NRQCD and tree-level SCET results can be
further investigated by the $J/\psi$ angular distribution. In Fig.
2(b), we show the angular coefficient $\alpha$ defined in Eq.
(\ref{cos}) as the function of $z$. The dashed, dotted and solid
curves are for the NRQCD, tree-level SCET and the interpolated
resummed results, respectively. The resummed SCET result has the
same $\alpha(z)$ as that of the tree-level SCET. $\alpha$ in the
NRQCD calculation is around zero in the region $z<0.85$, while falls
off rapidly as $z>0.85$. At $z=1$, $\alpha$ is about $-0.85$. In
contrast, $\alpha$ in the tree-level or resummed SCET result almost
does not change with $z$. This behavior provides some hint about why
the tree-level SCET result does not match with the NRQCD one very
well at large $z$. It was known that at the end point $z=1$, only
the scalar component of the gluon-gluon system is allowed, which
gives $\alpha=-0.85$ \cite{brodsky}. Apart from the end point, other
spin components should be involved and might give dominant
contributions which increase $\alpha$ fast to be around zero with
decreasing $z$. The leading SCET expansion in the small parameter
$\lambda\sim\sqrt{1-z}$, which gives rise to a scalar operator for
the gluon-gluon system (Eqs. (\ref{operator}) and
(\ref{coefficient})), cannot describe the contributions from other
spin components. This implies that the power counting rules of SCET
might break down due to some yet unknown reasons. One possibility is
that part of the power suppressed contributions might be
kinematically enhanced significantly. If this were true,  one would
have to match onto SCET to the next-to-leading order in $\lambda$,
obtain the power suppressed operators and their coefficients, and
then perform the resummation procedure. However this complicated
calculation goes far beyond the purpose of this paper.

It is well known that the scale of $J/\psi$ is a little awkward for
the application of NRQCD, thus one might worry whether the
discrepancy between the NRQCD and tree-level SCET results is just an
illusion. The key observation here is that, the only dimensionless
parameter in this process is $r=M^2/s$. Therefore the normalized
cross section $\sigma/R$, which is dimensionless, should only depend
on $r$. That means even in a model world in which $J/\psi$ could be
chosen to be very heavy (for example $30~\gev$), the $J/\psi$
momentum spectrum would still be the same as that showed in Fig. 2
if $r=M^2/s$ is taken to be fixed by increasing the c.m. energy
$\sqrt{s}$ correspondingly. That is to say, even in a model world in
which the application of NRQCD is guaranteed by very massive
$J/\psi$ and the perturbative treatment of jet function is
guaranteed by the larger c.m. energy $\sqrt{s}$, the large
discrepancy between the NRQCD and tree-level SCET calculations would
still be there.

As we have emphasized before the similarity between the radiative
$\Upsilon$ decay $\Upsilon \to \gamma gg$ and the inclusive $J/\psi$
production $\gamma^\ast \to J/\psi gg$, one might naturally ask whether
there is similar trouble for the former case. As shown in Ref. \cite{fleming},
for the photon momentum spectrum, there is no significant discrepancy
between the NRQCD and tree-level SCET results for the radiative
$\Upsilon$ decay. However Ref. \cite{fleming} has not investigated the angular
distribution of photons. Considering the process $e^+ e^- \to \Upsilon \to
\gamma gg$ at CLEO, $\Upsilon$ is transversely polarized in the c.m. frame.
Accordingly the ultrasoft matrix element, which is proportional to
$v^\delta v^{\delta'}-g^{\delta \delta'}$ in Ref. \cite{fleming}
($v$ is the $\Upsilon$ velocity), now should change to be proportional
to $-g^{\delta \delta'}_{\perp e}$ (Eq. (\ref{pola})) by
using vacuum-saturation approximation. Performing an analogous
calculation as what we have done in the last section, we obtain the
differential
$\Upsilon$ decay rate in the tree-level SCET
\begin{equation}\label{ucos}
\frac{d \Gamma_{\rm trSCET}}{\Gamma_0 dz d \cos \theta}
=\frac{3}{8}z( 1 + \cos^2 \theta),
\end{equation}
where $z=2E_\gamma/M_\Upsilon$ and $\theta$ is the scattering angle
between the momentum of the outgoing photon and the electron
beamline in the c.m. frame. Here $\Gamma_0$ is a normalization
constant. Eq. (\ref{ucos}) indicates that the angular coefficient
$\alpha(z)$ defined in Eq. (\ref{cos}) is always equal to unity at
any $z$ in the SCET calculation.

The resummed SCET momentum distribution is the same as that in
Ref. \cite{fleming}, while we choose the interpolation way as in
Eq. (\ref{int}) in order to give an interpolated resummed result
for both momentum and angular distribution.

In Fig. 3, we show the momentum distributions and the coefficient $\alpha$
of the radiative $\Upsilon$ decay as the function of $z$ in the NRQCD (the
dashed line), the tree-level SCET
(the dotted line), the resummed SCET (the dot-dashed line) and the
interpolated resummed calculations. The NRQCD result is taken from
Ref. \cite{Koller}. It is clear that although the momentum distribution
in the tree-level SCET calculation is close to that in NRQCD,
the large difference of $\alpha(z)$ still exists in the end point region.
Similar with the $J/\psi gg$ production, the gluon-gluon system here also
leaves only scalar component at the end point $z=1$ \cite{majp}.
Therefore the NLO matching onto SCET might also play an important role
in this process.

\begin{figure}
\centerline{\epsfysize 2.0 truein \epsfbox{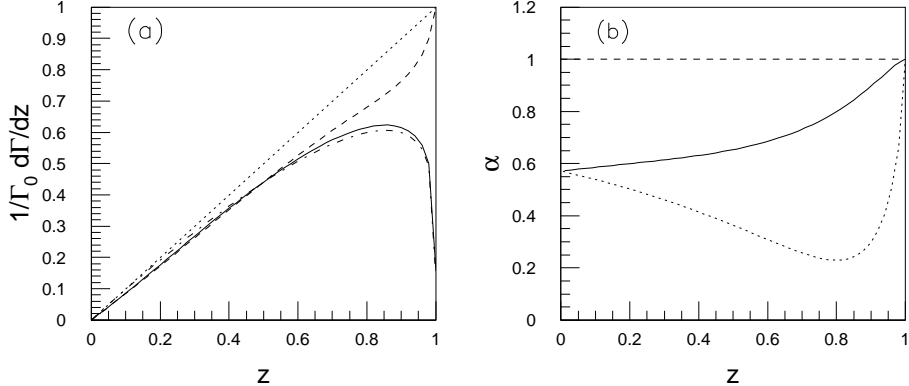}}
\caption{The momentum distribution (a) and angular coefficient $\alpha(z)$
(b) for $e^+ e^- \to \Upsilon \to \gamma gg$ process. The dashed, dotted
and dot-dashed curves correspond to the NRQCD, tree-level SCET and
resummed SCET calculations, respectively. The solid curve is for the
interpolated resummed results.}
\end{figure}

In this paper, we studied the collinear suppression effect in the
process $e^+e^-\to J/\psi gg$ and performed a leading power
calculation in SCET. We obtained the decreasing $J/\psi$ spectrum in
the endpoint region, which comes from
the Sudakov logarithms suppression, and combine our SCET result with
the NRQCD calculation. We then showed the momentum and angular distributions
for $J/\psi$ in SCET and compared with the NRQCD results. Surprisingly, we
found that, even before the resummation over large logarithms, there
already exists a large discrepancy between the SCET and NRQCD results in
the endpoint region of $J/\psi$ spectrum. A similar discrepancy is also
found in the angular distribution of the radiative $\Upsilon$ decay.
Therefore it should be
highly interesting to have further investigations, for example, including
the power suppressed contributions, on these processes.

\section*{Acknowledgments}
We would like to thank Kaoru Hagiwara and Deshan Yang for helpful
discussions. ZHL is supported by the Japan Society for the Promotion of
Science (JSPS).


\begin{thebibliography}{99}

\bibitem{NRQCD}
G.T. Bodwin, E. Braaten, and G.P. Lepage, Phys. Rev. D {\bf 51}, 1125
(1995) [Erratum-ibid. D {\bf 55}, 5853 (1997)].

\bibitem{CDF}
CDF Collaboration, F. Abe {\it et al.}, Phys. Rev. Lett.  {\bf 79}, 572
(1997); {\bf 79}, 578 (1997).

\bibitem{cho}
P. Cho and A.K. Leibovich, Phys. Rev. D {\bf 54}, 6690 (1996);
F. Yuan, C.F. Qiao, and K.T. Chao, Phys. Rev. D {\bf 56}, 321 (1997);
S. Baek, J. Lee, H.S. Song, and P. Ko, J. Korean Phys.i Soc. {\bf 33}, 97
(1998); S. Baek, P. Ko, J. Lee, and H.S. Song, hep-ph/9804455.

\bibitem{Lin}
K. Hagiwara, E. Kou, Z.-H. Lin, C.-F. Qiao, and G.-H. Zhu, hep-ph/0401246.

\bibitem{braaten-chen}
E. Braaten and Y.Q. Chen, Phys. Rev. Lett.  {\bf 76}, 730 (1993).

\bibitem{babar}
BaBar Collaboration, B. Aubert {\it et al.}, Phys. Rev. Lett. {\bf 87},
162002 (2001).

\bibitem{belle1}
Belle Collaboration, K. Abe {\it et al.}, Phys. Rev. Lett. {\bf 88},
052001 (2002); ibid. {\bf 89}, 142001 (2002)

\bibitem{belle2}
P. Pakhlov, {\it Measurement of double $c{\bar c}$ production},
the talk given on 22 - 29 March 2003, Les Arcs, France.

\bibitem{FLM03}
S. Fleming, A.K. Leibovich, and T. Mehen, Phys. Rev. D {\bf 68}, 094011
(2003).

\bibitem{pe-ope}
F. Maltoni and A. Petrelli, Phys. Rev. D {\bf 59}, 074006 (1999);
I.Z. Rothstein and M.B. Wise, Phys. Lett. B {\bf 402}, 346 (1997).

\bibitem{SCET}
C.W. Bauer {\it et al.}, Phys. Rev. D {\bf 63}, 014006 (2001);
C.W. Bauer {\it et al.}, Phys. Rev. D {\bf 63}, 114020 (2001);
C.W. Bauer and I.W. Stewart, Phys. Lett. B {\bf 516}, 134 (2001);
C.W. Bauer {\it et al.}, Phys. Rev. D {\bf 65}, 054022 (2002).

\bibitem{bauer}
C.W. Bauer {\it et al.}, Phys. Rev. D {\bf 64}, 114014 (2001).

\bibitem{fleming}
S. Fleming and A.K. Leibovich, Phys. Rev. Lett, {\bf 90}, 032001 (2003);
Phys. Rev. D{\bf 67}, 074035 (2003).

\bibitem{soto}
X. Garcia i Tormo and J. Soto, Phys. Rev. D{\bf 69}, 114006 (2004).

\bibitem{CLEO}
CLEO Collaboration, B. Nemati {\it et al.}, Phys. Rev. D {\bf 55}, 5273
(1997).

\bibitem{Keung}
W.Y. Keung, Phys. Rev. D {\bf 23}, 2072 (1981);
J.H. K\"{u}hn and H. Schneider, Phys. Rev. D {\bf 24}, 2996 (1981);
Z. Phys. C {\bf 11}, 263 (1981);
V.M. Driesen, J.H. K\"{u}hn, and E. Mirkes, Phys. Rev. D {\bf 49}, 3197 (1994).

\bibitem{brodsky}
S.J. Brodsky, A.S. Goldhaber, and J. Lee, Phys. Rev. Lett {\bf
91}, 112001 (2003); S. Dulat, K. Hagiwara, and Z.-H. Lin, hep-ph/040223.

\bibitem{Koller}
K. Koller and T. Walsh, Nucl. Phys. B {\bf 140}, 449 (1978).

\bibitem{majp}
X.-G. He, H.-Y. Jin, and J.P. Ma, Phys. Rev. D {\bf 66}, 074015 (2002).

\end{thebibliography}
\end{document}